\documentclass[10pt,a4paper]{iopart}

\pdfoutput=1

\usepackage{graphicx}
\usepackage{latexsym}
\usepackage{iopams}
\usepackage{bbm}
\usepackage[colorlinks,linkcolor=blue,anchorcolor=black,citecolor=red]{hyperref}

\newcommand{\N}{{\mathbb{N}}}  
\newcommand{\R}{{\mathbb{R}}}  

\newcommand{\feta}{\boldsymbol{\eta}}

\renewcommand{\P}{{\mathbb{P}}}

\newcommand{\be}{\begin{equation}}
\newcommand{\ee}{\end{equation}}
\newcommand{\bee}{\begin{equation*}}
\newcommand{\eee}{\end{equation*}}

\begin{document}

\title{Explosive condensation in symmetric mass transport models}

\author{Yu-Xi Chau$^1$, Colm Connaughton$^{1,2,3}$, Stefan Grosskinsky$^{1,2}$}

\address{$^1$ Centre for Complexity Science, University of Warwick, Coventry CV4 7AL, UK}
\address{$^2$ Mathematics Institute, University of Warwick, Coventry CV4 7AL, UK}
\address{$^3$ London Mathematical Laboratory,  14 Buckingham St., London WC2N 6DF, UK}

\ead{\mailto{y-x.chau@warwick.ac.uk}, \mailto{c.p.connaughton@warwick.ac.uk}, \mailto{s.w.grosskinsky@warwick.ac.uk}}

\begin{abstract}
We study the dynamics of condensation in a misanthrope process with nonlinear jump rates and factorized stationary states. For large enough density, it is known that such models have a phase separated state, with a non-zero fraction of the total mass concentrating in a single lattice site. It has been established in [B Waclaw and M R Evans, Phys.\ Rev.\ Lett., 108(7):070601, 2012] for asymmetric dynamics that such processes exhibit explosive condensation, where the time to reach the stationary state vanishes with increasing system size. This constitutes a spatially extended version of instantaneous gelation which has previously been studied only in mean-field coagulation models. We show that this phenomenon also occurs for symmetric dynamics in one dimension if the non-linearity is strong enough, and we find a coarsening regime where the time to stationarity diverges with the system size for weak non-linearity. In higher space dimensions explosive condensation is expected to be generic for all parameter 
values. Our results are based on heuristic mean field arguments which are confirmed by simulation data.
\end{abstract}

\maketitle

\section{Introduction}\label{sec:intro}


The statistical mechanics of non-equilibrium cluster growth is a topic of recurring scientific interest due to a wide range of applications in polymer physics, atmospheric science, soft matter and astrophysics (see \cite{krapivsky_kinetic_2010} and the references therein). A large collection of clusters is studied, each of which is an aggregate of constituent units called monomers. Clusters grow as a result of pairwise interactions. Two basic growth mechanisms are common in practice: coagulation and particle exchange. In coagulating systems, when a cluster of size $i$ interacts with a cluster of size $j$, a single cluster of size $i+j$ is formed. In particle exchange models, when a cluster of size $i$ interacts with a cluster of size $j$, a monomer is exchanged, resulting in a pair of clusters with sizes $i-1$ and $j+1$. Related is the well-known Becker--D\"{o}ring model of homogeneous nucleation where monomer exchange is indirect:  clusters grow or decay by absorbing or shedding one monomer at a time (see \cite{ball_becker-doring_1986} and the references therein). In both coagulation and particle exchange models, the rates of interaction generally depend on the sizes, $i$ and $j$, of the interacting clusters. This dependence is encoded in a (usually symmetric) function, $K(i,j)$, 
of the cluster sizes called the interaction kernel. Given $K(i,j)$, one fundamental question of practical importance is to understand the rate at which large clusters are formed. 

A common feature of both processes which is relevant to this question is a dynamical transition known as gelation. It occurs when the kernel is a sufficiently rapidly increasing function of the particle sizes. Gelation was originally discovered as a singularity in the mean-field coagulation equations resulting in the divergence of the second moment of the cluster size 
distribution in a finite time, $t^*$. For a full discussion see \cite{leyvraz_scaling_2003} and the references therein. The mean field equations for exchange driven growth can exhibit similar behaviour \cite{ben-naim_exchange-driven_2003}. The meaning of this singularity becomes clear if one considers a finite Markov chain \cite{aldous_deterministic_1999} containing $N$ particles and studies its behaviour in the limit $N\to \infty$. This  was done explicitly for coagulation with the product kernel, $K(i,j) = ij$, in \cite{lushnikov_exact_2005}. It was shown that, for fixed $N$, a finite fraction of the total mass in the system is collected into a single "super-particle" within a time $t_N$. As $N\to \infty$, $t_N$ tends to a finite value, $t^*$, known as the gelation time and the mass contained in the super-particle diverges.  Gelation in an infinite system is therefore interpreted as the formation of a cluster of arbitrarily large size within a finite time.

Soon after the discovery of the gelation transition, it was conjectured \cite{spouge_monte_1985,van_dongen_possible_1987} that for some kernels, the gelation time, $t^*$, is zero. The counter-intuitive idea that arbitrarily large clusters could form in arbitrarily short time is called instantaneous gelation.  This effect was proven rigorously for coagulation models in \cite{jeon_spouges_1999}. A similar regime  is observed for exchange-driven growth \cite{ben-naim_exchange-driven_2003} and for the Becker-D\"{o}ring model \cite{brilliantov_nonscaling_1991,king_asymptotic_2002}. Since it exhibits a singularity at time 0+, instantaneous gelation cannot be described using mean field kinetic equations without introducing some kind of regularisation. Such regularisation could be done by introducing an explicit cut-off into the kinetic theory as was done for coagulation in \cite{ball_instantaneous_2011} or by heuristically incorporating a finite system size, $N$, into the mean field equations as was done for exchange-driven growth in \cite{ben-naim_exchange-driven_2003}. In both cases, there is 
strong evidence that while the gel time indeed decreases as the regularisation is removed, this decrease is logarithmically slow. Consequently it is expected to be difficult to  distinguish between regular finite time gelation and instantaneous gelation in practice.

Until recently, it was widely believed that instantaneous gelation was a pathological property of mean-field kinetic equations and could not occur in spatially extended systems. A counter-example has recently been found \cite{Waclaw:2012ww}  in a variation of a spatially extended exchange-driven growth model of misanthrope type. Misanthrope processes \cite{misanthrope,Evans:2004} are lattice models in which particles jump between neighbouring sites at rates which are a function of the number of particles at both the origin and destination sites. If one thinks of each site on the lattice as a cluster, the analogy with exchange-driven growth becomes clear. Such processes have been extensively studied because they can exhibit stationary measures which can be factorised as a product of single site measures and calculated analytically. Furthermore, these stationary measures can exhibit an interesting transition as the total particle density is varied.  If the density exceeds a critical value, a finite 
fraction of all particles concentrates on a single lattice site, whereas the bulk of the system is distributed homogeneously at the critical density. This phenomenon is referred to as condensation and the site on which a fraction of all particles concentrates is called the condensate. The connection between condensation and gelation emerges when one considers the dynamics of how the condensate is formed in time. The condensate is the analogue of the super-particle discussed above. The question of how the condensate forms dynamically cannot be answered solely from knowledge of the stationary product measure itself. The model studied in 
\cite{Waclaw:2012ww} was found to generate a condensate in a time which decreases logarithmically with the system size, like in the case of instantaneous gelation. They called this effect "explosive condensation". 

The dynamics of condensation in spatially homogeneous particle systems has already been studied heuristically in \cite{Drouffe:1999jk} and in subsequent work on zero-range processes \cite{Evans:2000vw,Godreche:2003gb,Grosskinsky:2003eq,Godreche:2005kj,Evans:2005fz} and related models \cite{Godreche:2001vo,Godreche:2007gx}. In these processes the jump rate of particles depends only on the occupation number on the departure site, and condensation occurs if this is a sufficiently fast decreasing function. There is also a significant literature on the dynamics of condensation in spatially heterogeneous models (see \cite{Godreche:2012ky} and references therein) which is not discussed here. Mathematically rigorous results in the homogeneous case have first been obtained for stationary distributions (see \cite{Jeon:2000wq,Grosskinsky:2003eq,Armendariz:2008cp} and references therein), and more recently also on the stationary dynamics of the condensate (see e.g.\ \cite{Beltran:2011cg,
agl2}), and the formation of the condensate from homogeneous initial conditions \cite{Beltran:2015}. The hydrodynamic behaviour has been studied heuristically in \cite{Schutz:2007ec}. In condensing zero-range type models the motion of large clusters is the slowest time scale, exchange of particles occurs through fluctuations in size, and relaxation time scales diverge with the system size.

While misanthrope-type processes share the same factorized stationary states as zero-range processes, the dynamics of condensation is entirely different \cite{Chleboun:2013um, Evans:2014dy}. Large clusters move on the same time scale as they exchange particles, or merge when they meet. This has first been established for the inclusion process on a rigorous level in \cite{Grosskinsky:2013ji} and with heuristic extensions in \cite{caoetal14}. This process has been introduced in \cite{Giardina:2007em} and is basically a linear version of the explosive condensation model, where condensation occurs only due to the scaling of a system parameter \cite{Grosskinsky:2011uh}. This induces a separation of time scales, and relaxation times diverge with the system size. Besides applications to energy transport \cite{Giardina:2009jx}, the inclusion process can also be interpreted as a multi-allele version of the Moran model \cite{MORAN:1962uk}. The model studied here would be a non-linear extension of the latter, where the 
competitive advantage of a group increases super-linearly with the group size. Further applications are discussed in \cite{Evans:2014dy} and references therein. 

\begin{figure}
\begin{center}
\mbox{\includegraphics[width=0.42\textwidth]{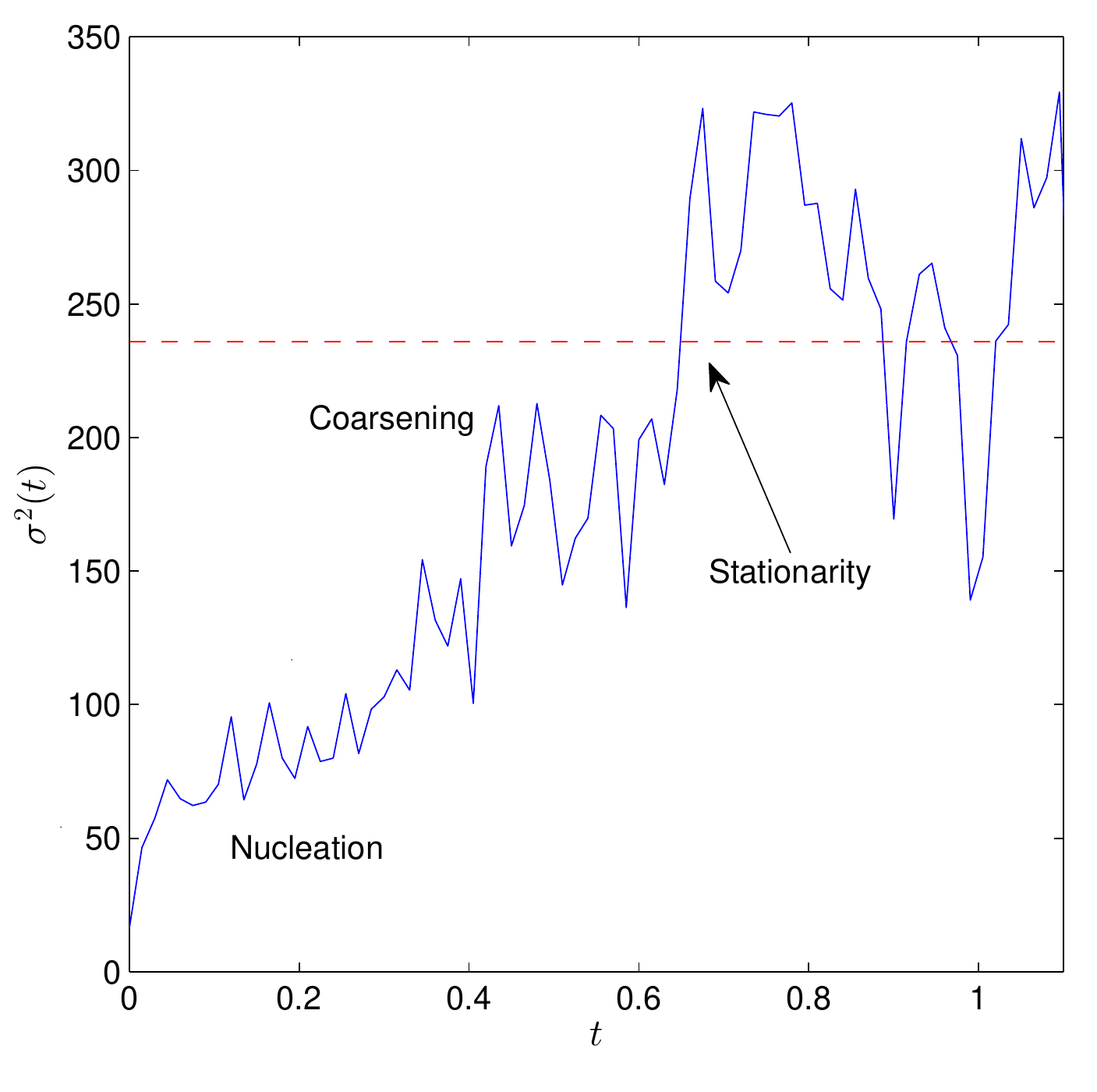}
\includegraphics[width=0.58\textwidth]{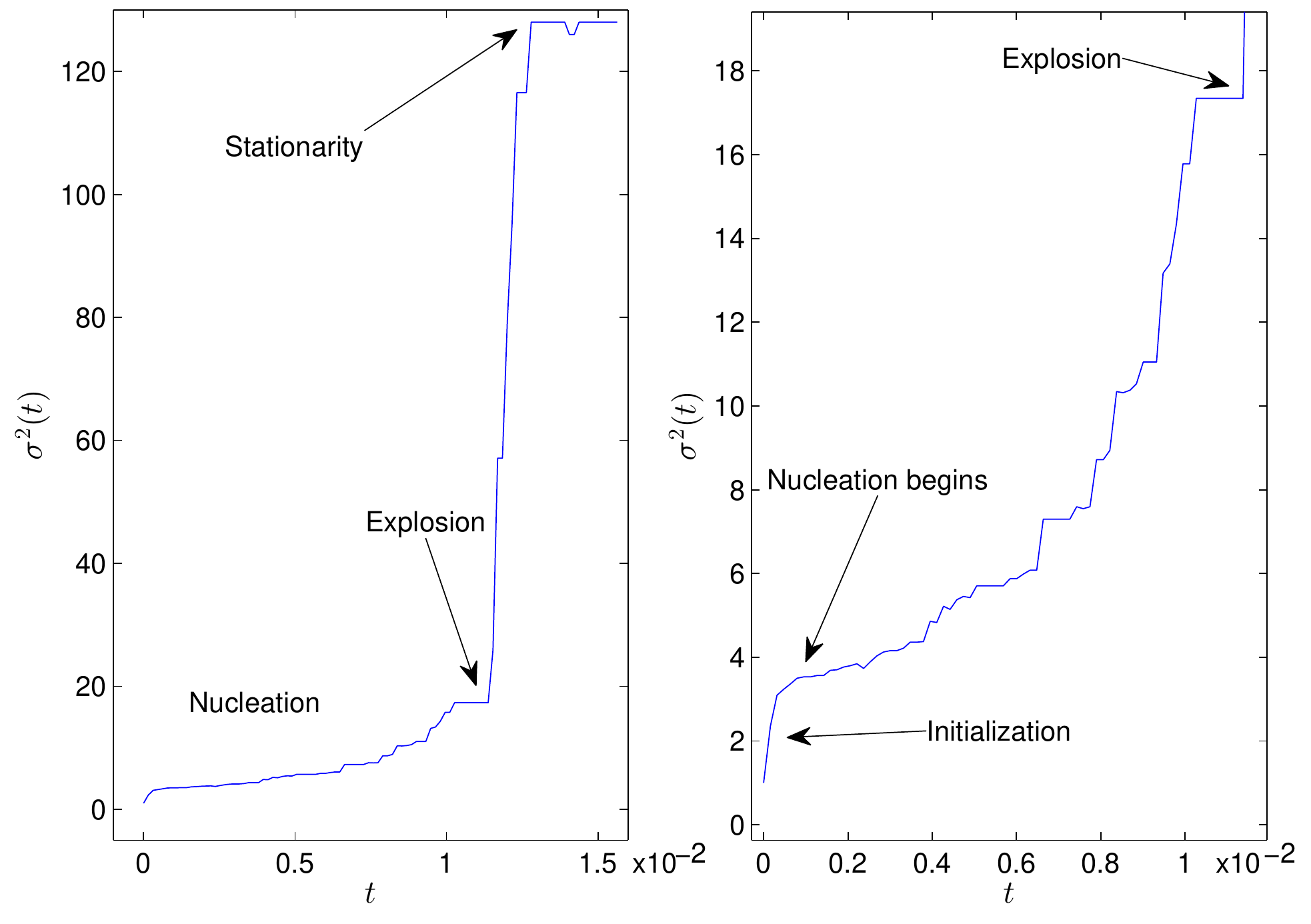}}
\end{center}
\caption{\label{fig:sketch}
The two different mechanisms of condensation, with coarsening (left) for $\gamma =2.25$ and explosion (middle and right) for $\gamma =4.5$, are illustrated using the second moment of occupation numbers $\sigma^2$ (\ref{sigmas}). For explosion, nucleation of clusters leads to a gradual increase (zoom on right) until one cluster reaches a critical size. In contrast, the coarsening regime (left) is not well separated from nucleation and both lead to a gradual increase to reach stationarity. Further parameter values are $L=128$ and $\rho =4$, $d=0.2$ (left) and $\rho =1$, $d=0.3$ (right).
}
\end{figure}

The model of \cite{Waclaw:2012ww} was completely asymmetric in the sense that particles could only move to the right. A natural question to ask is whether the results on explosive condensation 
depend crucially on the asymmetry of the dynamics which facilitates mass transport. In this paper we show that this is not the case, and establish a detailed picture of the condensation dynamics for a symmetric system on a one-dimensional lattice with periodic boundary conditions. If the non-linearity of the jump rates is strong enough (characterized by a system parameter $\gamma >3$) explosive condensation occurs analogously to the asymmetric system. Local fluctuations of the initial conditions are amplified and clusters form in an initial nucleation regime. As soon as one of them reaches a critical size explosion occurs, the cluster moves fast enough to cover the whole lattice in a vanishing amount of time and grow into the condensate. For weaker non-linearity (with system parameter $\gamma\in (2,3)$), the critical cluster size cannot be reached and several macroscopic clusters form during nucleation. 
These are then moving across the lattice, exchanging mass in a coarsening regime very similar to the dynamics in inclusion processes, and the time to reach a single condensate diverges with the system size. These two mechanisms are illustrated in Figure \ref{fig:sketch} using the second moment of the occupation numbers (cf.\ (\ref{sigmas})) as a suitable observable to characterize the distance to stationarity. The second mechanism does not exist for asymmetric dynamics or symmetric dynamics in higher dimensions. Our derivation also applies to asymmetric systems and all dimensions, and provides a slight simplification of the arguments presented in \cite{Waclaw:2012ww}.

The paper is organised as follows. In Section \ref{sec:model} we introduce the model and summarize known results on the stationary distributions, emphasizing differences to zero-range models regarding the stationary particle current or activity. In Section \ref{sec:cond} we present our main results on the different stages and mechanisms of condensation dynamics. We derive the critical cluster size as a key concept, and provide heuristic arguments supported by simulation data for explosive and non-explosive condensation mechanisms. In Section \ref{sec:sum} we conclude with a summary and a short discussion of expected results in higher space dimensions.

\section{The model}\label{sec:model}

\subsection{Definition and notation}\label{sec:ip:def}

The explosive condensation model $(\feta (t):t\geq 0)$ is a stochastic particle system defined on a lattice $\Lambda_{L}$ of $L$ sites, which we fix to be one-dimensional with periodic boundary conditions. Configurations are denoted by $\feta =(\eta_{x}:x\in\Lambda_{L})\in X_{L}$ where $\eta_{x}\in\N$ is the number of particles at site $x\in\Lambda_{L}$. This can be arbitrarily large and the full state space is given by $X_{L}=\N^{\Lambda_{L}}$.

A particle at site $x$ jumps to site $y\neq x$ with rate
\be\label{rates}
c(\feta ,x,y):=p(x,y) \eta_x^\gamma (d+\eta_y^\gamma )\quad\mbox{with parameters}\quad \gamma >1 ,d>0\ .
\ee
The purely spatial part $p(x,y)\geq 0$ is given by rates of an irreducible random walk on the lattice $\Lambda$, which determines the dynamics of a single particle together with the parameter $d>0$. In this paper we focus on symmetric nearest neighbour dynamics with
\be
p(x,y)=\frac12 \delta_{y,x+1} +\frac12 \delta_{y,x-1}\ ,
\ee
other choices are shortly discussed in Section \ref{sec:sum}. The interaction part of the rates is such that larger occupation numbers on the departure or target site increase the jump rate in a super-linear fashion, controlled by the parameter $\gamma >1$. These rates are a slight variation of the ones introduced in \cite{Waclaw:2012ww}, which are 
\be\label{warates}
\big((\eta_x +d)^\gamma -d^\gamma \big)(d+\eta_y )^\gamma \ .
\ee
We present our analysis and numerical results for the rates (\ref{rates}) since they have a slightly simpler form. For $\gamma =1$ both rates coincide, and correspond to the previously studied inclusion process \cite{Grosskinsky:2013ji,caoetal14}. They show the same asymptotic behaviour and all results, including factorized steady states as well as condensation dynamics, hold in an analogous fashion in both models. For reference, if spatial structure is neglected and $d$ is small compared to site occupation numbers, the rate equations for this model are equivalent to those studied for exchange-driven growth by Ben-Main and Krapivsky \cite{ben-naim_exchange-driven_2003} with the simple product kernel $K(i,j) = (i\,j)^\gamma$. For reasons, explained below, we shall be interested in the regime $\gamma>2$. At the mean field level, this regime should exhibit instantaneous gelation according to the analysis of  \cite{ben-naim_exchange-driven_2003}.

The dynamics of the continuous-time process $(\feta (t):t\geq 0)$ is then given by the usual master equation
\be
\frac{d}{dt} p_t [\feta ]=\sum_{x,y\in\Lambda} \Big( c(\feta^{y,x} ;x,y) p_t [\feta^{y,x} ]-  c(\feta ;x,y)\, p_t [\feta ] \Big)
\ee
for all $\feta\in X_L$, with the shorthand $p_t [\feta ]=\P [\feta_t =\feta ]$. We use the notation $\feta^{y,x}$ with $\eta^{y,x}_z =\eta_z -\delta_{z,y} +\delta_{z,x}$ for the configuration $\feta$ with one particle moved from site $y$ to $x$, and the usual convention that $p_t [\feta ]=0$ if $\eta_z <0$ for some $z\in\Lambda$.

\subsection{Stationary distributions}\label{sec:ip:stat}

Factorized stationary states for this class of misanthrope processes are well established, see \cite{Evans:2004} as well as \cite{Chleboun:2013um,Evans:2014dy} and references therein. Since we focus on translation invariant systems, we have homogeneous factorized distributions
\be
\nu_{\phi}^{L}[\feta]=\prod_{x\in\Lambda_{L}}\nu_{\phi}[\eta_{x}]\, \quad\textrm{with marginals}\quad \nu_{\phi}[n]=\frac{1}{z(\phi)}w(n)\phi^{n} \ .
\ee
Here the stationary weights are given by
\be\label{weights}
w(n)=\prod_{k=1}^n \frac{(k-1)^\gamma +d}{k^\gamma} \sim n^{-\gamma } \quad\mbox{as }n\to\infty\ ,
\ee
where the asymptotic behaviour is the same as for the rates (\ref{warates}) used in \cite{Waclaw:2012ww}. So the single-site partition function is
\be
z(\phi)=\sum_{k=0}^{\infty}w(k)\phi^{k}<\infty \quad\mbox{for all }\phi\in [0,1]\ ,
\label{partition}
\ee
since we assume $\gamma >1$. The fugacity parameter $\phi$ controls the average particle density in the grand-canonical setting, which is given by the expectation
\be
R(\phi):=\langle \eta_{x}\rangle_\phi =\sum_{k=0}^{\infty}k\nu_{\phi}[k]=\phi\partial_{\phi}\log_{z}(\phi)\ .
\label{avgden}
\ee
This is a monotone increasing function with $R(0)=0$ and maximal value $\rho_c :=R(1)\in (0,\infty ]$. 
In complete analogy to previous results on zero-range processes \cite{Drouffe:1999jk,Evans:2000vw,Godreche:2003gb,Grosskinsky:2003eq} and the model with rates (\ref{warates}) in \cite{Waclaw:2012ww}, the process exhibits condensation in the thermodynamic limit if the critical density $\rho_c <\infty$, which is the case if and only if $\gamma >2$. In that case, the system phase separates into a homogeneous background with density $\rho_c$, and the excess mass (of order $(\rho -\rho_c )L$ particles) concentrates on a single lattice site, called the condensate.

Mathematically, this is formulated in terms of the canonical distributions
\be\label{candis}
\pi_{L,N} [\feta ]:=\frac{\mathbf{1}_{X_{L,N}} (\feta )}{Z_{L,N}} \prod_{x\in\Lambda_L} w(\eta_x )\ ,
\ee
which concentrate on configurations $X_{L,N} =\big\{ \feta\in X_L :\sum_{x\in\Lambda_L } \eta_x =N\big\}$ with a fixed particle number $N\geq 0$, and the normalization $Z_{L,N}$ is simply given by a finite sum over all stationary weights. The process is ergodic on the finite set $X_{L,N}$, and $\pi_{L,N}$ is the unique stationary distribution. Condensation can then be understood in terms of the equivalence of canonical and grand-canonical ensembles in the thermodynamic limit with density $\rho\geq 0$,
\be\label{equi}
\pi_{L,N} \to\left\{\begin{array}{cl} \nu_\phi ,&\ R(\phi )=\rho \leq\rho_c \\ \nu_1 ,&\ \rho\geq \rho_c \end{array}\right.\ ,\quad\mbox{as }N,L\to\infty ,\ N/L\to \rho\ .
\ee
This has been established rigorously in \cite{Jeon:2000wq,Grosskinsky:2003eq,Armendariz:2008cp,Armendariz:2013} and holds in a weak sense, i.e. canonical expectations $\langle f\rangle_{L,N}$ of bounded, local functions $f:X_{L,N} \to\R$ converge to the corresponding expectation $\langle f\rangle_\phi$ under the grand-canonical product distribution. For $\rho <\rho_c$ this also holds for unbounded integrable functions $f$ \cite{Chleboun:2013um}.

The interpretation is that for $\rho >\rho_c$ the system exhibits a phase separated (condensed) state with a single condensate as explained above.
Depending on the initial condition, the condensed phase can have larger initial volume which shrinks during time evolution, as is the case in the well-studied coarsening process for zero-range dynamics \cite{Godreche:2005kj,Grosskinsky:2003eq,Beltran:2015} or inclusion processes \cite{Grosskinsky:2013ji,caoetal14}. The homogeneous (or bulk) phase of the system is distributed with a truncated version of the critical distribution $\bar\nu_1$, where occupation numbers are bounded by a size of order $L$ due to the presence of the condensed phase. This detail will become important later to compute expectations in a phase separated state.

\subsection{Fundamental diagram}\label{sec:ip:fd}

\begin{figure}
\begin{center}
\mbox{\includegraphics[width=0.45\textwidth]{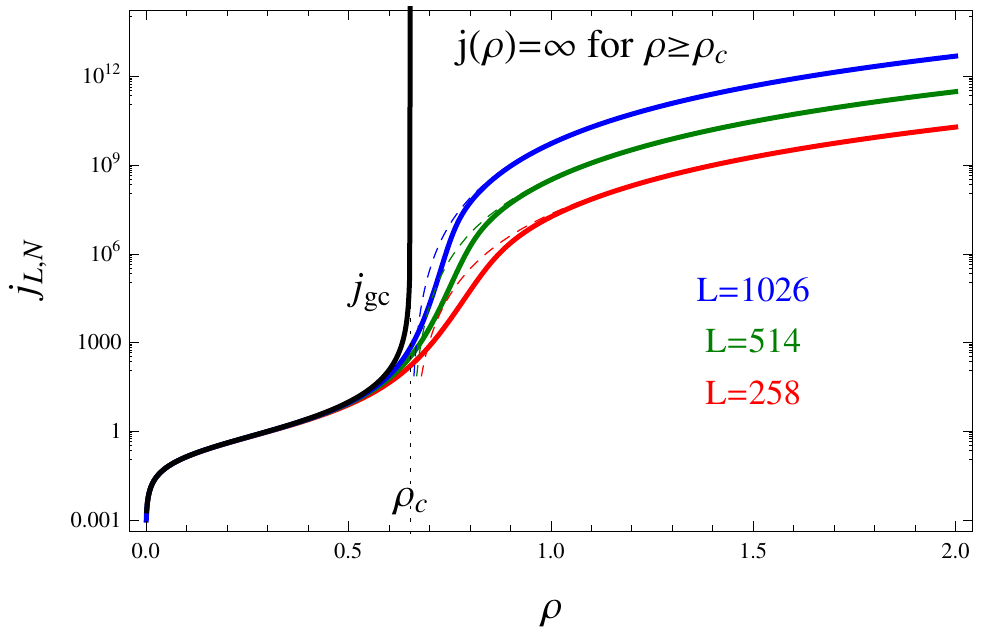}\quad\includegraphics[width=0.45\textwidth]{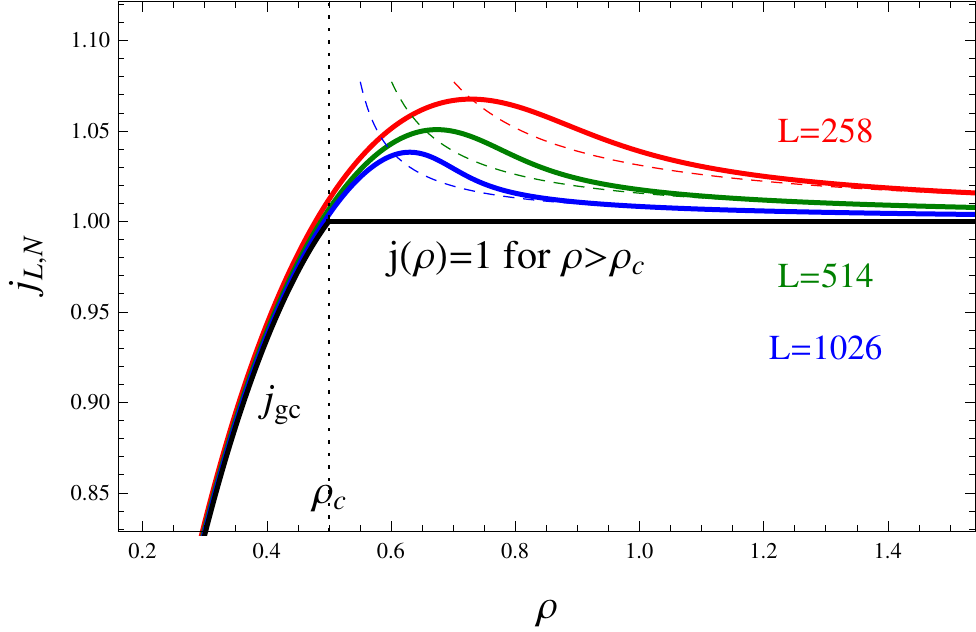}}
\end{center}
\caption{\label{fig:jln}
Canonical current $j_{L,N}$ (\ref{cancurr}) as a function of density $\rho =N/L$ for different system sizes $L$ (full coloured lines) for the explosive condensation model with $d=1$ and $\gamma =5$ (left) and for a zero-range process with rates $g(n)=1+b/n$ with $b=4$ (right), resulting from exact recursions in the canonical ensemble (see e.g.\ \cite{chlebounetal10}). It is approximated by the grand-canonical current $j_{gc}$ (\ref{gccurr}) for $\rho <\rho_c$ (full black line) 
and by (\ref{jlnapp}) for supercritical densities in the explosive model (dashed lines). The finite-size corrections in zero-range processes are also given by the current in the condensed phase as explained in the text. Note that both plots have very different scales.
}
\end{figure}

A typical stationary configuration in the explosive condensation model looks almost identical to that of extensively studied condensing zero-range processes with rates $g(n)=1+b/n$, since the stationary weights $w(n)$ have the same power-law decay. 
However, as has already been observed in \cite{Waclaw:2012ww} for totally asymmetric rates, the dynamics of the process is very different. On the level of stationary distributions this is signified by the expected jump rate
\be\label{cancurr}
j_{L,N} :=\big\langle \eta_x^\gamma (d+\eta_y^\gamma )\big\rangle_{L,N}\ ,
\ee
and its behaviour in the thermodynamic limit
\be\label{fund}
j(\rho ):=\lim_{N,L\to\infty\atop N/L\to\rho} j_{L,N} \ .
\ee
This corresponds to the activity in a symmetric system, and to the stationary current in an asymmetric system where $j(\rho )$ determines the fundamental diagram of the model. For simplicity we refer to this quantity simply as current in the following. With the equivalence of ensembles, the limiting current can be computed at least for $\rho <\rho_c$ under the factorized grand-canonical distribution choosing $\phi$ according to (\ref{equi}),
\be\label{gccurr}
j_{gc} (\phi ):=\big\langle \eta_x^\gamma (d+\eta_y^\gamma )\big\rangle_\phi =\langle \eta_x^\gamma \rangle_\phi \big( d+\langle \eta_y^\gamma \rangle_\phi \big)\ .
\ee
For general zero-range processes it is well known that $j_{gc} (\phi )=\phi$, and therefore $j(\rho )$ is simply given by the inverse of (\ref{avgden}) for $\rho <\rho_c$. Since the rates $g(n)=1+b/n$ are bounded, (\ref{equi}) also implies that $j(\rho )=1$ for all $\rho\geq \rho_c$. The convergence for finite $L$ is dominated by the jump rate $g((\rho -\rho_c )L)$ out of the condensate \cite{chlebounetal10} (as shown in Figure \ref{fig:jln}). So the condensed phase in zero-range models does not contribute to the current in this model, and it is known that the condensate remains at a fixed position and only moves on a very slow timescale \cite{Godreche:2005kj,Beltran:2011cg,agl2}.

For the model studied here, however, the rates are not bounded and it is clear from the power-law decay in (\ref{weights}) and (\ref{gccurr}) that the current diverges as $\rho\to\rho_c$. This indicates that the condensed phase contributes to, and in fact dominates the stationary current for super-critical systems. As opposed to zero-range processes, the condensate moves with a high rate of order $L^\gamma$ as is discussed in more detail later. In the condensed state of a system with $N$ particles on $L$ sites where $N/L>\rho_c$, the canonical stationary current is therefore asymptotically given by
\be
j_{L,N} \simeq\frac{2}{L} (\rho -\rho_c )^\gamma L^\gamma \langle \overline{\eta_x^\gamma }\rangle_1 +\frac{L-1}{L} \langle\overline{ \eta_x^\gamma }\rangle_1^2 \ .
\ee
The first contribution is due to the condensate, and the second is a bulk contribution determined by expectations w.r.t.\ the truncated critical distribution $\bar\nu_1$, and we omitted lower order terms involving the parameter $d$. Note that $\langle \eta_x^\gamma \rangle_1 =\infty$, but in the phase separated state bulk occupation numbers are bounded by the condensate size of order $(\rho -\rho_c )L$, and the truncated moment is therefore also $\langle \overline{\eta_x^\gamma }\rangle_1 \leq (\rho -\rho_c )L$. Thus
\be\label{jlnapp}
j_{L,N} \simeq 2(\rho -\rho_c )^{\gamma +1} L^{\gamma} +\mathrm{const.}\ L^2\ ,
\ee
which diverges as $L\to\infty$, and since $\gamma >2$ the current is dominated by the condensed phase. The different behaviour for the fundamental diagrams of zero-range and explosive condensation models is illustrated in Figure \ref{fig:jln}.

The most striking dynamic effect of this model, however, is the possibility of explosive condensation, i.e.\ the condensed stationary state is reached in a time that decreases with the system size $L$. This has been established in \cite{Waclaw:2012ww} for totally asymmetric dynamics for all $\gamma >2$. While there is no a-priori reason to expect mean field theory to be relevant to dynamics on a one-dimensional lattice, it is interesting to remark that the explosive condensation regime, $\gamma>2$, corresponds with the instantaneous gelation regime at mean-field level. In the following we investigate this phenomenon for symmetric mass transport dynamics.

\section{Main results on condensation dynamics\label{sec:cond}}

In contrast to the totally asymmetric system, which exhibits explosive condensation for all $\gamma >2$, the symmetric model only shows explosion for $\gamma >3$ and the time to stationarity diverges for $\gamma \in (2,3)$. In the latter regime the approach to stationarity is dominated by a cluster coarsening dynamics similar to inculsion processes \cite{Grosskinsky:2013ji,caoetal14}. The main difference between the two scenarios is whether the largest cluster is able to reach a critical size $m_c$ which has to scale sublinearly with the system size $L$, as is discussed in the next subsection. Once it exceeds that size, it will move across the lattice in a time vanishing with system size, and the model exhibits explosion. Both cases can be distinguished by a very different scaling of the expected time to stationarity $\langle T_{SS}\rangle$. This can be defined as
\be\label{tss}
T_{SS}:=\inf\big\{ t>0 :\max_{x\in\Lambda} \eta_x (t)\geq (\rho -\rho_c )L\big\}\ ,
\ee
i.e. the time for the process to reach a fully condensed configuration with a condensate of size $(\rho -\rho_c)L$.

\subsection{Nucleation and critical cluster size\label{sec:nuc}}

The role of the initial conditions is not crucial for our result, as long as they are homogeneous. We use two such conditions, for the first one we simply place all $N$ particles independently on a uniformly chosen site. This leads to a multinomial distribution for $\feta$, and fluctuations of the maximal occupation number are of order $\log L$. We also confirmed our results numerically for deterministic initial conditions where each site contains exactly $\rho$ particles for integer densities. This has no fluctuations and naturally leads to slower dynamics than the first choice. 
But also in this case the fluctuations in the dynamics lead to clusters of size $\log L$ within arbitrarily small time intervals, since the dynamics is driven by $L$ independent Poisson processes. The resulting equilibration dynamics for both initial conditions shows the same scaling and only differs in prefactors. 

\begin{figure}
\begin{center}
\mbox{\includegraphics[width=0.45\textwidth]{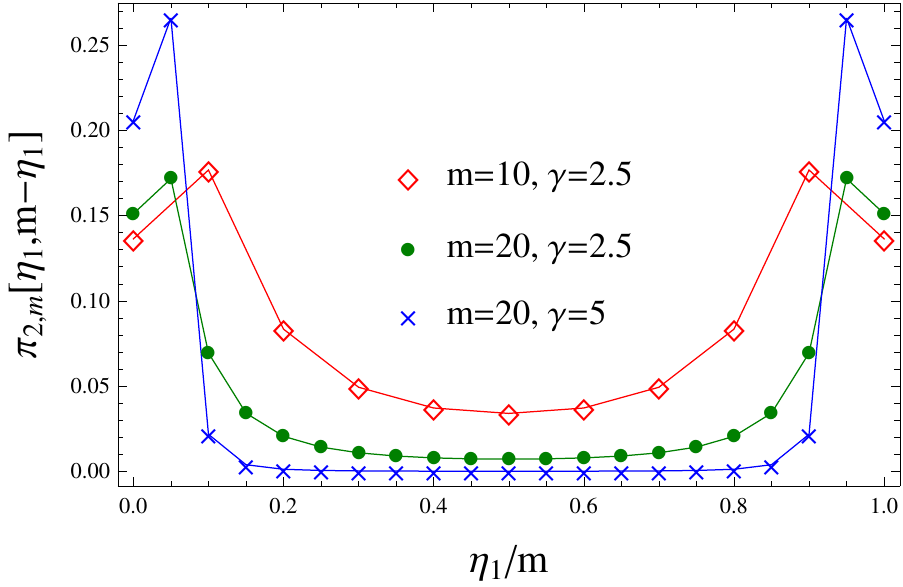}\quad\includegraphics[width=0.45\textwidth]{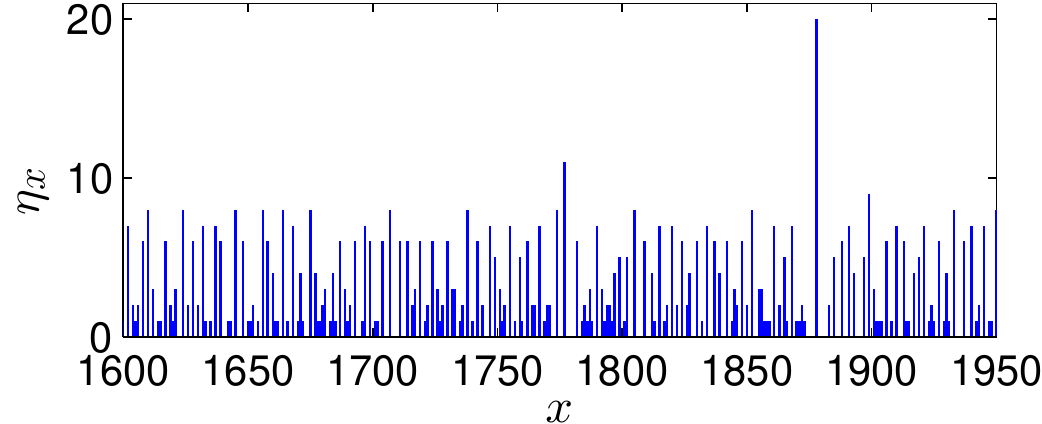}}
\end{center}
\caption{\label{fig:nucl}
Amplification of fluctuations through local dynamics. Left: the bimodality of the two-site distribution (\ref{bimodal}) increases with occupation numbers $m$ and the parameter $\gamma$ ($d=1$). This leads to rough configurations with increasing fluctuations as shown on the right for $\gamma =7$ ($d=1$, $\rho =2$, $L=2048$) at a very early time $3.56\times 10^{-7}$, just before one of the clusters starts dominating.
}
\end{figure}

The initial fluctuations in occupation numbers are amplified by the local dynamics and lead to the nucleation of larger clusters. Consider a two-site system with rates (\ref{rates}) and a total of $m$ particles, where $\eta_2 = m-\eta_1$. With (\ref{candis}) the stationary distribution is given by
\be\label{bimodal}
\pi_{2,m} [\eta_1 ,m-\eta_1 ]= \frac{w(\eta_1 )w(m-\eta_1 )}{Z_{2,m}}\ .
\ee
Since the weights $w$ have a power-law tail, the distribution is bimodal with the boundaries dominating more and more with increasing $m$ and increasing parameter $\gamma$ as shown in Figure \ref{fig:nucl}. 
So the mass between any two neighbouring sites tends to concentrate in one of them, which leads to rough configurations and the formation of clusters as dynamically stable objects.

These clusters then perform fluctuation activated motion, where a cluster at site $x$ jumps to a neighbouring site $y=x\pm 1$ with rate
\be\label{dm}
D(m)=m^\gamma (d+\langle \eta_y^\gamma \rangle )/m \sim m^{\gamma -1}\ .
\ee
Here we have used that the environment outside the cluster is not yet stationary, so $\langle \eta_y^\gamma \rangle =O(1)$ depending only on the particle density. So the cluster loses a particle at rate $O(m^\gamma )$, and the factor $1/m$ is the probability that the remaining particles will follow and the whole cluster will move to site $y$. This results from the probability that a symmetric random walk started in site $1$ hits site $m$ before site $0$, and this walk represents the distribution of mass among both sites corresponding to a path from one boundary to the other in Figure \ref{fig:nucl} (left). The walk is symmetric but continuous-time with large jump rates for roughly equal mass distribution, which leads to the bimodal structure of the stationary distribution. This also implies that the time for a cluster to step is dominated by the time it takes for the first particles to move, while all the others follow within the same time scale. Omitting terms involving the parameter $d$, we have to leading order for the corresponding time scale
\be
\sum_{k=1}^{m-1} \frac{1}{k^\gamma (m-k)^\gamma} \sim m^{1-2\gamma}\int_{1/m}^{1/2} x^{-\gamma}\, dx\sim m^{-\gamma}\ll m^{1-\gamma}\ ,
\ee
which justifies (\ref{dm}).

Since $\gamma >2$ jump rates of clusters increase super-linearly in their size, and through their symmetric motion clusters gain mass at rate $D(m)/m=m^{\gamma -2}$. Therefore the time it takes to grow to a size $m_L \gg \log L$ which is larger than the initial fluctuations, is proportional to
\be\label{nutime}
\int_{\log L}^{m_L}\!\! m^{2-\gamma}\, dm=\frac{m_L^{3-\gamma} {-}(\log L)^{3-\gamma} }{3-\gamma } \sim \left\{\begin{array}{cl}\! \frac{m_L^{3-\gamma}}{3-\gamma } &,\ \gamma\in (2,3)\\ \frac{(\log L)^{3-\gamma}}{\gamma -3} &,\ \gamma >3
\end{array}\right. \! .
\ee
This vanishes as $\gamma >3$ and the time to grow a cluster of any size is in fact dominated by the maximal initial size $\log L$. For $\gamma\in (2,3)$ the time diverges depending on the final size of the cluster $m_L$, and for macroscopic clusters of size $O(L)$ it scales like $L^{3-\gamma}$.

In this context, we define the critical cluster size $m_c$ as
\be\label{mc}
L^2 /D(m_c ) =L^2 /m_c^{\gamma -1} \sim 1\quad\Rightarrow\quad m_c \sim L^{2/(\gamma -1)}\ .
\ee
By definition, if $m\gg m_c$ the cluster visits all $L$ sites of the lattice instantaneously, i.e. in a time $o(1)$ as $L\to\infty$. Together with (\ref{nutime}) the system then exhibits explosive condensation, where equilibration is dominated by the fast motion of a single largest cluster. This is only possible if
\be
m_c \ll L\quad\Leftrightarrow\quad \gamma >3\ ,
\ee
whereas for $\gamma\in (2,3)$ we have $m_c \gg L$ and the system cannot exhibit explosion. We discuss both cases in detail below. Note that the nucleation of clusters we just discussed is dominated by purely local dynamics, the clusters then start to interact globally leading to equilibration of the system.

\subsection{Coarsening dynamics for $\gamma\in (2,3)$}

If $\gamma <3$ the critical cluster size $m_c \gg L$ (\ref{mc}) cannot be reached, and instead several clusters of macroscopic size $m=O(L)$ nucleate from the initial condition within a diverging time of order $L^{3-\gamma}$. The system reaches a phase separated state where the condensed phase is split in several isolated clusters, which move across the lattice and exchange mass in the global part of the dynamics. In this case equilibration is therefore dominated by a coarsening process of clusters similar to inclusion processes. We analyze this in a mean field approach close to the methods in \cite{caoetal14}, where we denote by $m(t)$ the average cluster size and by $n(t)$ the number of clusters at time $t$, such that by conservation of mass
\be\label{conse}
m(t)\, n(t)=(\rho -\rho_c )L\ .
\ee
Since the background in the phase separated state is stationary, we have to use the truncation $\langle \overline{\eta_y^\gamma }\rangle_1 \leq m$ in analogy with (\ref{jlnapp}), to get for the jump rate of a cluster
\be\label{dm2}
D(m)=m^\gamma (d+\langle \overline{\eta_y^\gamma }\rangle )/m \sim m^\gamma\ .
\ee
Note that this is larger than in the nucleation regime (\ref{dm}) due to enhanced fluctuations of the bulk. The average distance between clusters is
\be
s(t)=L/n(t)=m(t)/(\rho -\rho_c)\ ,
\ee
so they meet at a rate proportional to
\be\label{meetrate}
D(m(t))/s(t)^2 =m^{\gamma -2} (\rho -\rho_c )^2\ .
\ee
Since the dynamics is symmetric, when two clusters of size $m_1$ and $m_2$ meet they merge with probability proportional to $1/(m_1 +m_2)$, in analogy with results for the inclusion process \cite{Grosskinsky:2013ji,caoetal14}. The clusters start exchanging particles as soon as they are separated by only one site, and the occupation number of this intermediate site is a symmetric random walk in continuous time on the state space $\{ 0,\ldots ,m_1 +m_2\}$. If the clusters do not merge they still exchange a certain number of particles, which happens more frequently and leads to a mass exchange on the same order as merge events. This is explained in detail in \cite{caoetal14} for inclusion processes which correspond to $\gamma =1$, and applies similarly in the model studied here.

\begin{figure}
\begin{center}
\mbox{\includegraphics[width=0.95\textwidth]{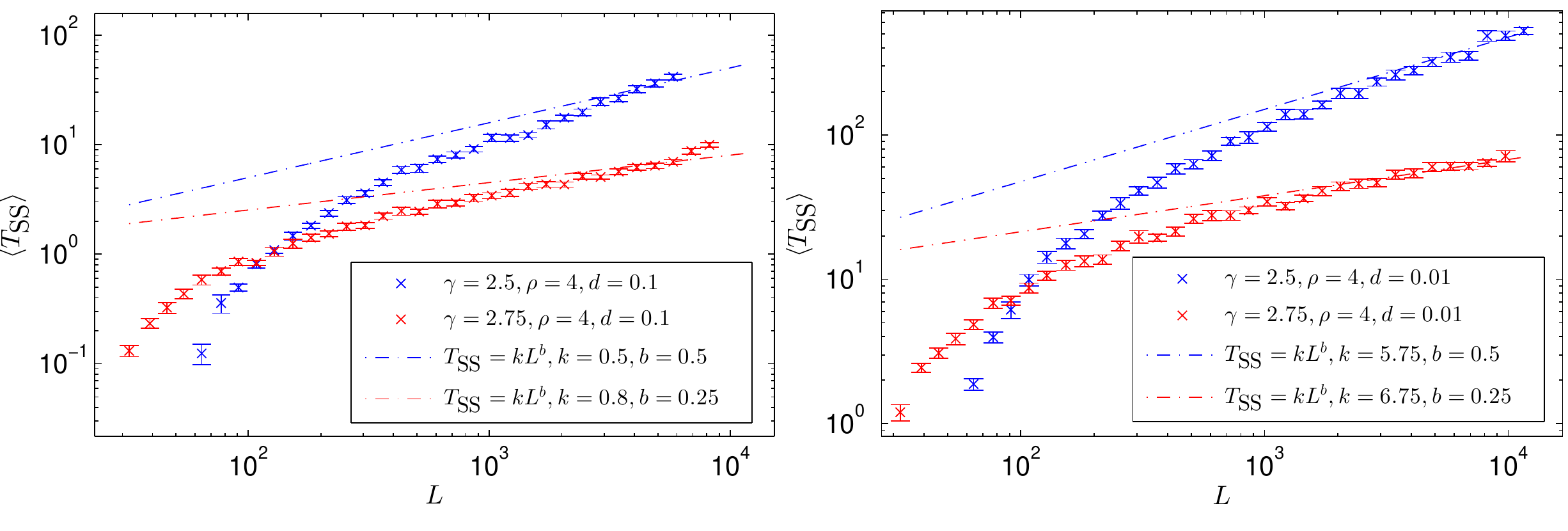}}
\end{center}
\caption{\label{fig:tss}
The expected time to stationarity $\langle T_{SS}\rangle$ (\ref{tss}) for $\gamma\in (2,3)$ increases with the system size, shown for different values of $\gamma$ and $d$. The predicted scaling law (\ref{tssc}) with fitted constants is denoted by dashed lines, data are averaged over $100$ realizations.
}
\end{figure}

If two clusters merge, the average size increases by
\be\label{change}
\Delta m(t)=\frac{(\rho -\rho_c )L}{n(t)-1}-\frac{(\rho -\rho_c )L}{n(t)} =\frac{m(t)}{n(t)-1}\sim m(t)\ .
\ee
Putting (\ref{meetrate}), (\ref{change}) and the merge probability of order $1/m(t)$ together, the change in cluster size is given by
\be\label{ccase}
\frac{d}{dt} m(t)=\mathrm{const.}\ m(t)^{\gamma -2}\ .
\ee
The solution with initial condition $m(0)\sim L$ is given by
\be\label{slaw}
m(t) =C (\rho -\rho_c )\, (t+t_0)^{1/(3-\gamma )}\quad\mbox{with}\quad t_0 =\frac{m(0)^{3-\gamma}}{C (\rho -\rho_c )}\ .
\ee
We pull out a factor $(\rho -\rho_c )$ from the multiplicative constant since simulations support the intuition that $n(t)$ is independent of the density, and $m(t)$ scales as $\rho -\rho_c$. 

\begin{figure}
\begin{center}
\mbox{\includegraphics[width=\textwidth]{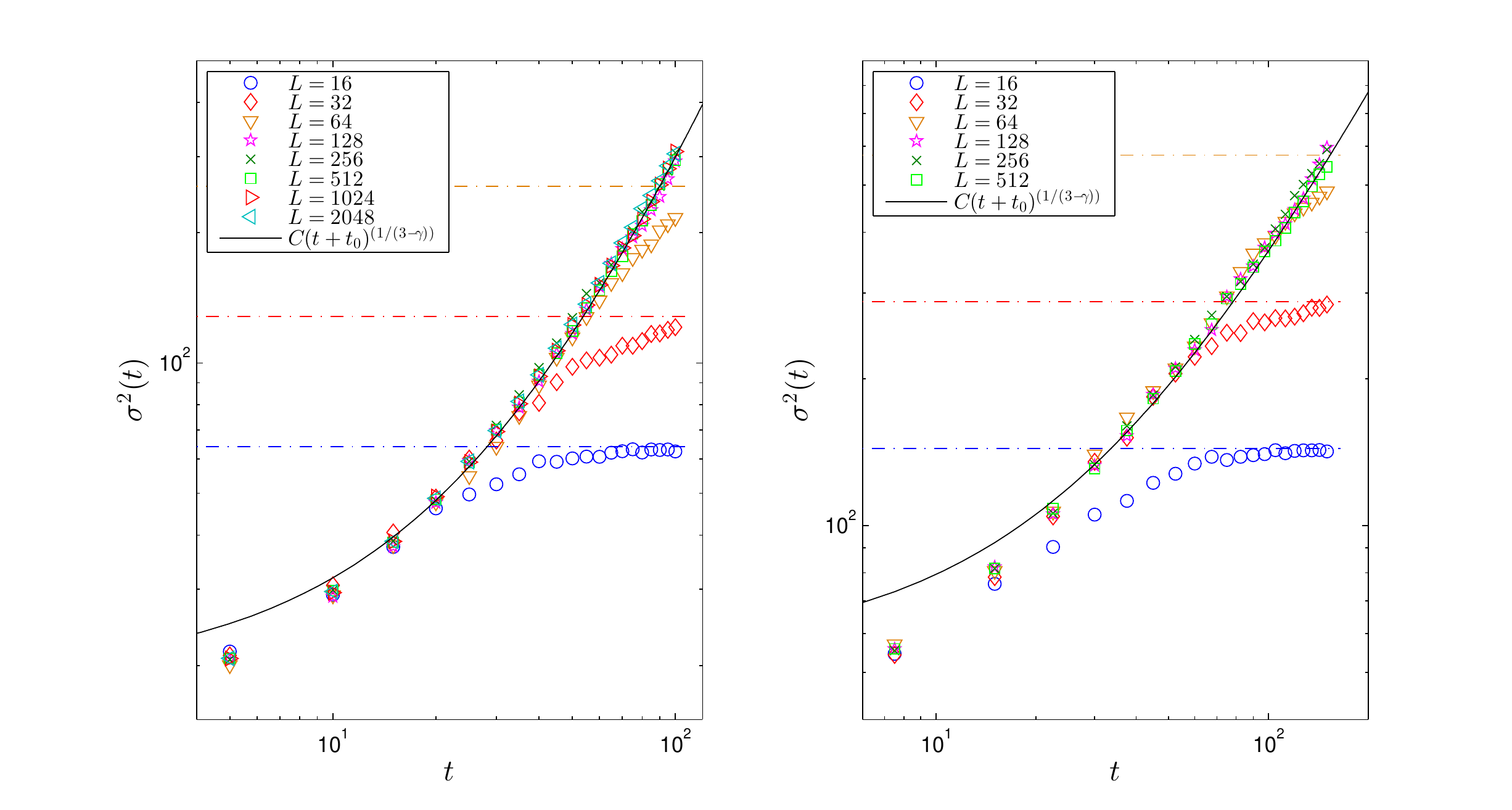}
}
\end{center}
\caption{\label{fig:scale}
The second moment $\sigma^2 (t)$ increases asymptotically as a power law predicted in (\ref{sconm}) given by a full line, before it saturates at the stationary value $(\rho -\rho_c )^2 L$ indicated by dashed lines. Parameter values are $\gamma =2.5$, $d=0.1$, $\rho =2$ (left), and $\gamma =2.25$, $d=0.01$, $\rho =3$ (right). Data points are averaged over $100$ realizations and errors are comparable to the size of the symbols. Fit constants are $t_0 =33.6$, $C=0.0168$ (left) and $t_0 =31.9$ and $C=0.545$(right), where $C$ includes the factor $(\rho -\rho_c )^2$ in (\ref{sconm}).
}
\end{figure}

From the scaling law (\ref{slaw}) it follows that the time to reach the stationary value $(\rho -\rho_c )L$ during coarsening scales like $L^{3-\gamma}$. 
This is of the same order as the time scale (\ref{nutime}) to grow clusters of macroscopic size, so nucleation and coarsening are not well separated in this model (cf.\ Figure \ref{fig:sketch}), in contrast to previous results for inclusion processes where separation results from the scaling of the system parameter $d$ \cite{Grosskinsky:2013ji,caoetal14}. In total, the time to stationarity scales like
\be\label{tssc}
\langle T_{SS} \rangle\sim L^{3-\gamma}\to\infty\quad\mbox{for }\gamma\in(2,3)\ ,
\ee
which is confirmed by simulation data presented in Figure \ref{fig:tss}.

To compare the scaling law (\ref{slaw}) to numerical data, we use the second moment as an observable which is numerically better accessible than the mean cluster size. In a phase separated state with $n(t)$ clusters of typical size $m(t)$ this is given by
\be\label{sigmas}
\sigma^2 (t):=\Big\langle\frac{1}{L}\sum_x \eta_x^2 (t)\Big\rangle =\frac{L-n(t)}{L} \langle \overline{\eta_x^2 }\rangle_1 +\frac{n(t)}{L}\, m^2 (t)\ .
\ee
Again, $\langle \eta_x^2 \rangle_1 =\infty$ for the critical product distribution with $\gamma\in (2,3)$, but using the truncation by $m=O(L)$ we see that this term is of order $L^{3-\gamma} \ll L$. So the bulk contribution can be neglected for large $L$, and with (\ref{conse}) we get
\be\label{sconm}
\sigma^2 (t)=(\rho -\rho_c) m(t)=C(\rho -\rho_c)^2 (t+t_0)^{1/(3-\gamma )}\ .
\ee
This is plotted in Figure \ref{fig:scale} and shows good agreement with simulation data in a coarsening window, before it saturates at its stationary value $(\rho -\rho_c )^2 L$.

\subsection{Explosive condensation for $\gamma >3$}

\begin{figure}
\begin{center}
\mbox{\includegraphics[width=0.95\textwidth]{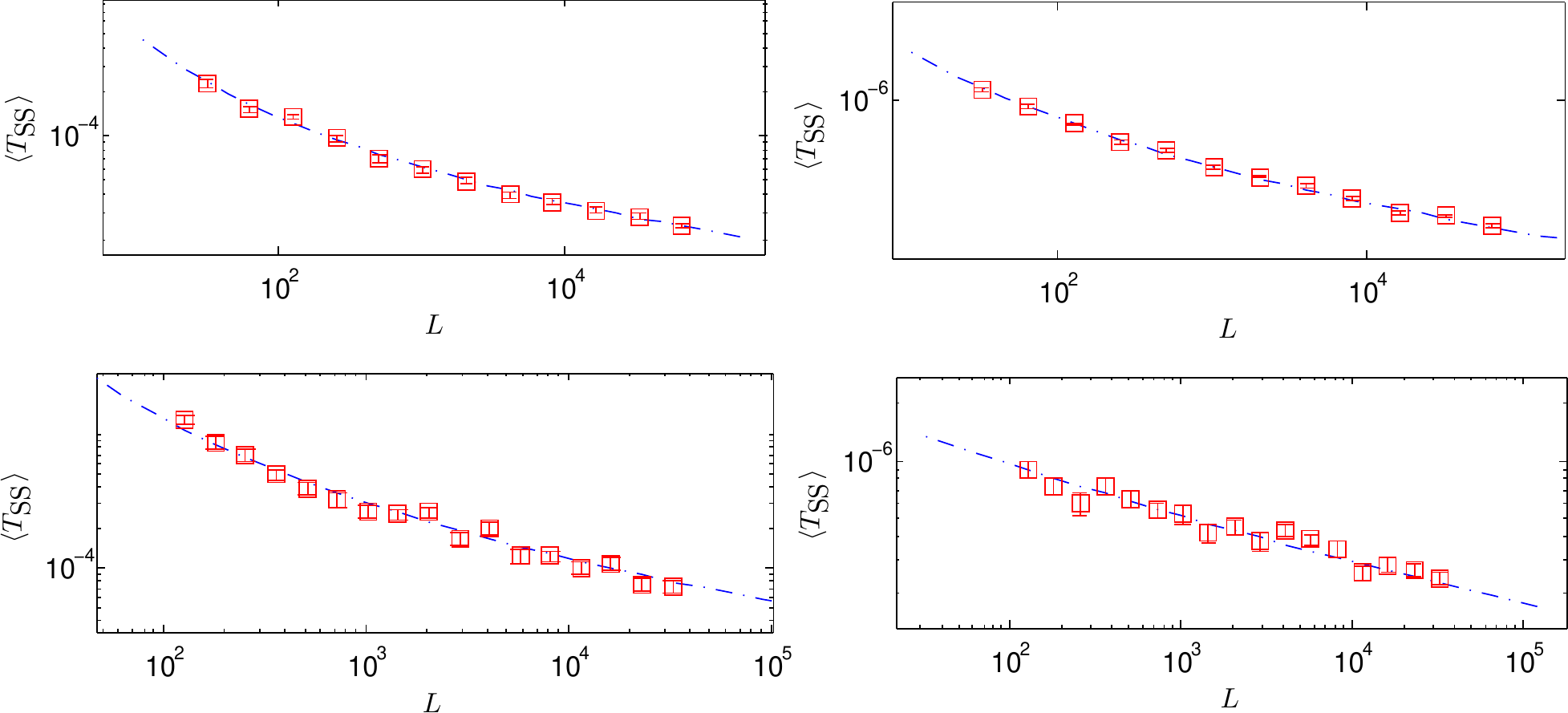}}
\end{center}
\caption{\label{fig:tss2}
The expected time to stationarity $\langle T_{SS}\rangle$ (\ref{tss}) for $\gamma >3$ decreases with the system size, shown for $\rho =2$ and different values $\gamma =5$, $d=0.1$ and $1$ (left), and $\gamma =7$, $d=0.1$ and $1$ (right). The predicted scaling law (\ref{tsse}) with a fitted constant is shown by blue dashed lines, data are averaged over $200$ realizations and errors are of the size of the symbols. 
}
\end{figure}

For $\gamma >3$ the critical cluster size $m_c \sim L^{2/(\gamma -1)}$ (\ref{mc}) scales sublinearly with $L$ and can be reached by the largest cluster in vanishing time of order $(\log L)^{3-\gamma}$ as discussed in (\ref{nutime}). From then on the largest cluster dominates the dynamics, visits all sites of the lattice and grows to a condensate containing all excess mass in a time vanishing with the system size $L$. 
As the speed of the clusters increases with increasing size, the time to stationarity is dominated by the nucleation time scale and we expect
\be\label{tsse}
\langle T_{SS}\rangle \sim (\log L)^{3-\gamma}\to 0\quad\mbox{for }\gamma >3\ .
\ee
This is plotted in Figure \ref{fig:tss2} and shows good agreement with simulation data.

\begin{figure}
\begin{center}
\mbox{\includegraphics[width=\textwidth]{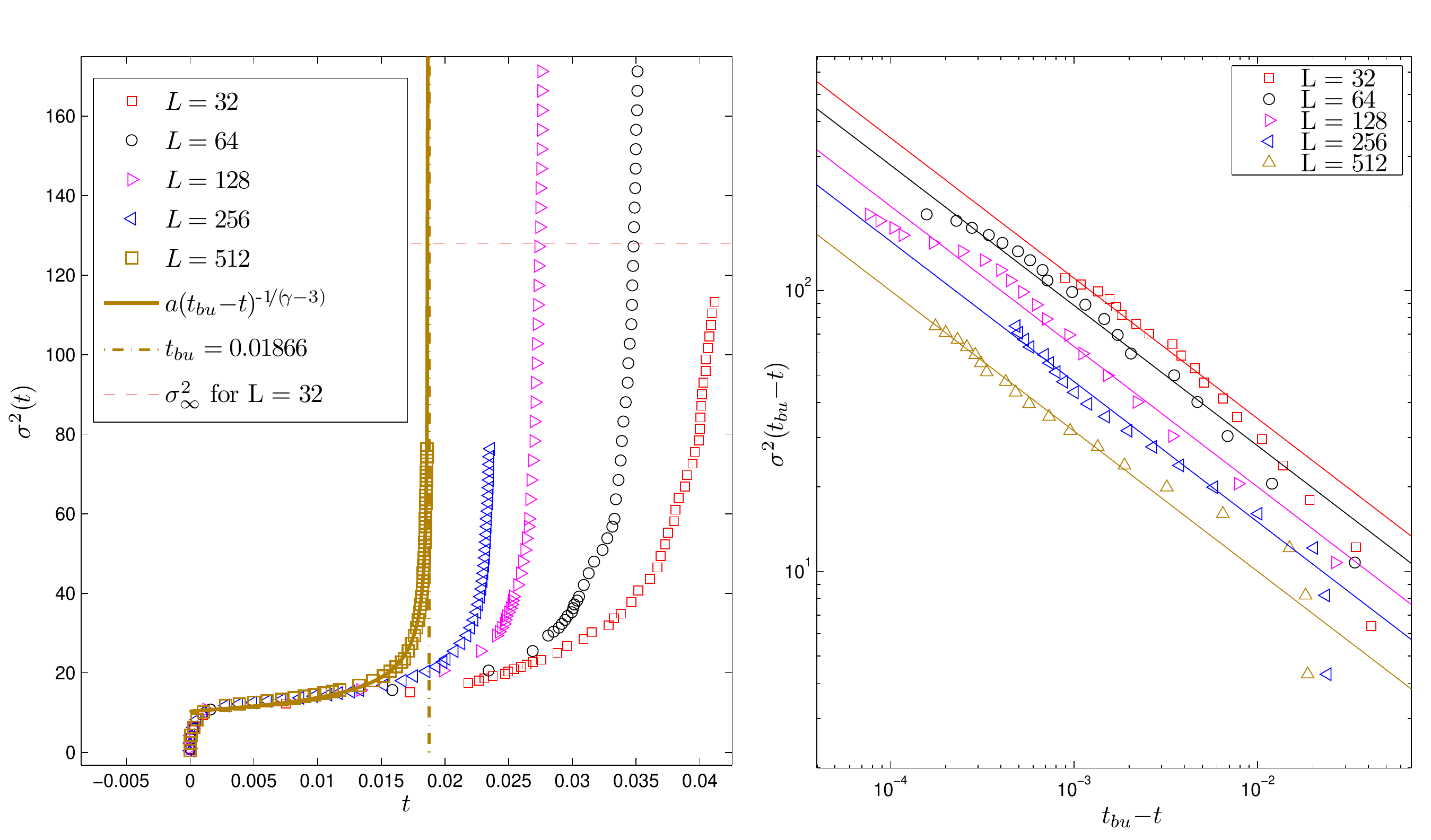}}
\end{center}
\caption{\label{fig:blowup}
The second moment $\sigma^2 (t)$ follows the prediction from (\ref{slaw2}) indicated by a full gold line for $L=512$ with corresponding fitted blow up time given by a dash dotted line (left). It saturates at the stationary value $(\rho -\rho_c )^2 L$ indicated by a dashed red line for $L=32$. Plotting against $(t_{bu}-t)$ in double-log scale confirms the functional behaviour following from (\ref{slaw2}) indicated by straight lines (right). Parameter values are $\gamma =5$, $\rho =2$ and $d=0.1$. Data points are averaged according to (\ref{tave}) over $50$ realizations and errors are of the size of the symbols.
}
\end{figure}

The time evolution of the average cluster size in this regime is dominated by the largest cluster, which gains mass at rate $D(m)/m$ as explained above. Since the bulk is still non-stationary we can use (\ref{dm}) for the cluster jump rate, and this leads to
\be\label{ecase}
\frac{d}{dt} m(t)=\mathrm{const.}\ m(t)^{\gamma -2}\ .
\ee
This turns out to coincide with (\ref{ccase}) for $\gamma <3$, even though the dynamics in both regimes look very different. For $\gamma >3$ the right-hand side grows super-linearly in $m$ which leads to a finite-time blowup of the solution,
\be\label{slaw2}
m(t) =C(\rho -\rho_c )\, (t_{bu} -t)^{-1/(\gamma -3)}\quad\mbox{with}\quad t_{bu} =\frac{m(0)^{3-\gamma}}{C(\rho -\rho_c )}\ .
\ee
The blow-up time $t_{bu}$ is determined by the initial condition $m(0)$, and since $m(0)\sim\log L$ we see that
\be
t_{bu}\sim (\log L)^{3-\gamma}\to 0\quad\mbox{as }L\to\infty\ .
\ee
This is consistent with explosive condensation and the scaling of $\langle T_{SS}\rangle$ in (\ref{tsse}). Using again the connection $\sigma^2 (t)=(\rho -\rho_c )m(t)$ analogous to (\ref{sconm}), we confirm (\ref{slaw2}) by comparison with numerical data in Figure \ref{fig:blowup}.

The explosive dynamics in this regime is completely analogous to the totally asymmetric case studied in \cite{Waclaw:2012ww}. Our derivation of the scaling of $\langle T_{SS}\rangle$ can be directly adapted to that case, and provides an alternative and slightly simpler version of the arguments presented there.

\begin{figure}
\begin{center}
\mbox{\includegraphics[width=\textwidth]{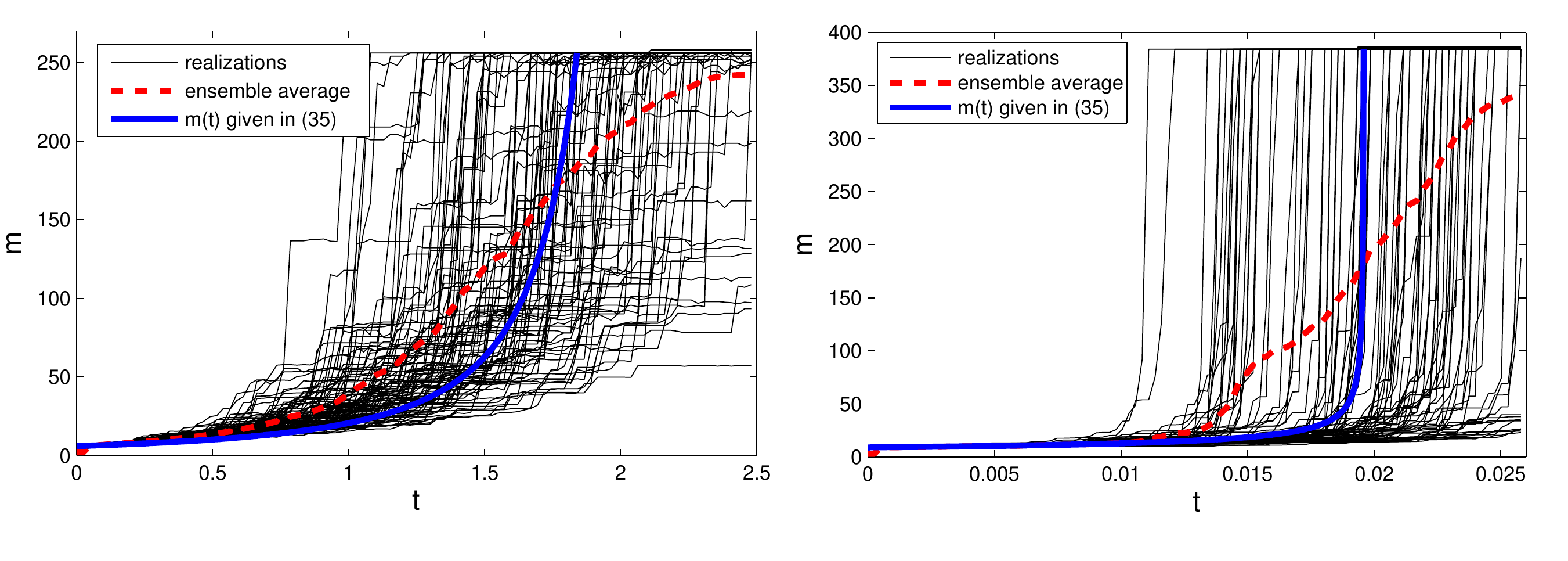}}
\end{center}
\caption{\label{fig:average}
Individual trajectories for the largest cluster size shown in black compare well with the prediction (\ref{slaw2}) shown as a thick blue curve, modulo a random time shift. The usual ensemble average (dashed red curve) does not coincide with the typical behaviour, and representative averages of these dynamics are therefore given by (\ref{tave}), which is used in Figure \ref{fig:blowup}. Parameter values are $L=128$, $d=0.1$, and $\gamma =3.5$, $\rho =2$ (left), and $\gamma =5$, $\rho =3$ (right). For larger $\gamma$, fluctuations of individual trajectories get smaller.
}
\end{figure}

Note that in contrast to the case $\gamma <3$, the nucleation and explosive condensation regime are well separated, and a transition occurs when the largest cluster reaches the critical size $m_c$, as can also be seen in Figure \ref{fig:sketch}. 
Since the nucleation regime is dominated by local dynamics, the time to reach $m_c$ is random and as discussed in Section \ref{sec:nuc} is basically determined by a maximum of $L$ Poisson random variables. The fluctuations of the latter are on the same scale $\log L$ as the expectation, leading to persistent randomness even with increasing system size $L$. 
Once $m_c$ is reached and the explosion mechanism takes over, relevant observables $\sigma^2 (t)$ and the size of the largest cluster $m(t)$ follow an almost deterministic trajectory with very small fluctuations (see Figure \ref{fig:average}). Variations in system trajectories therefore result from a random time shift corresponding to completion of the nucleation regime, and after that typical individual trajectories $\frac{1}{L}\sum_x \eta_x (t)^2$ follow (\ref{slaw2}) where $t_{bu}$ is replaced by a random time (see blue curve in Figure \ref{fig:average}). This implies that the typical behaviour is very different from the ensemble average $\sigma^2 (t)$ (red dashed curve in Figure \ref{fig:average}). Instead of averaging at fixed times $t$, it is thus more informative to average the time it takes different trajectories to reach a level $\sigma^2$ of the second moment, i.e.
\be\label{tave}
t(\sigma^2 ):=\Big\langle\inf\big\{ s>0:\frac{1}{L}\sum_x \eta_x (t)^2\geq \sigma^2 \big\}\Big\rangle\ .
\ee
Inverting this provides the behaviour of a typical trajectory of the system which is used to average the data in Figure \ref{fig:blowup}. 

This behaviour is in contrast to the case $\gamma <3$, where typical trajectories fluctuate around the mean with fluctuations decreasing with the system size, due to self-averaging effects arising from global coarsening dynamics. This results in a decreasing error with $L$ in Figure \ref{fig:tss}, whereas it is largely independent of $L$ in Figure \ref{fig:tss2}.

\subsection{The boundary case $\gamma =3$}

\begin{figure}
\begin{center}
\mbox{\includegraphics[width=0.6\textwidth]{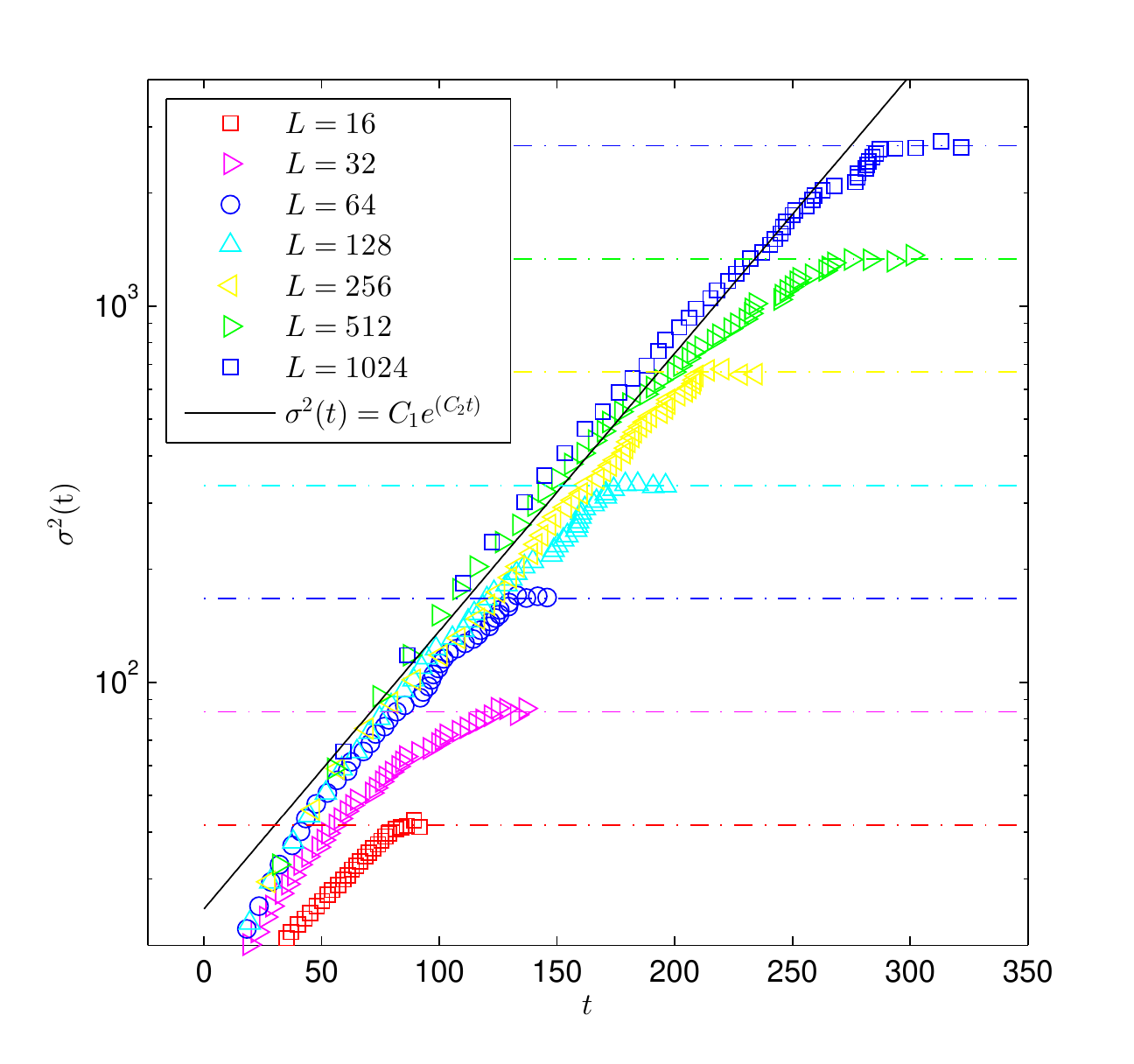}
\includegraphics[width=0.4\textwidth]{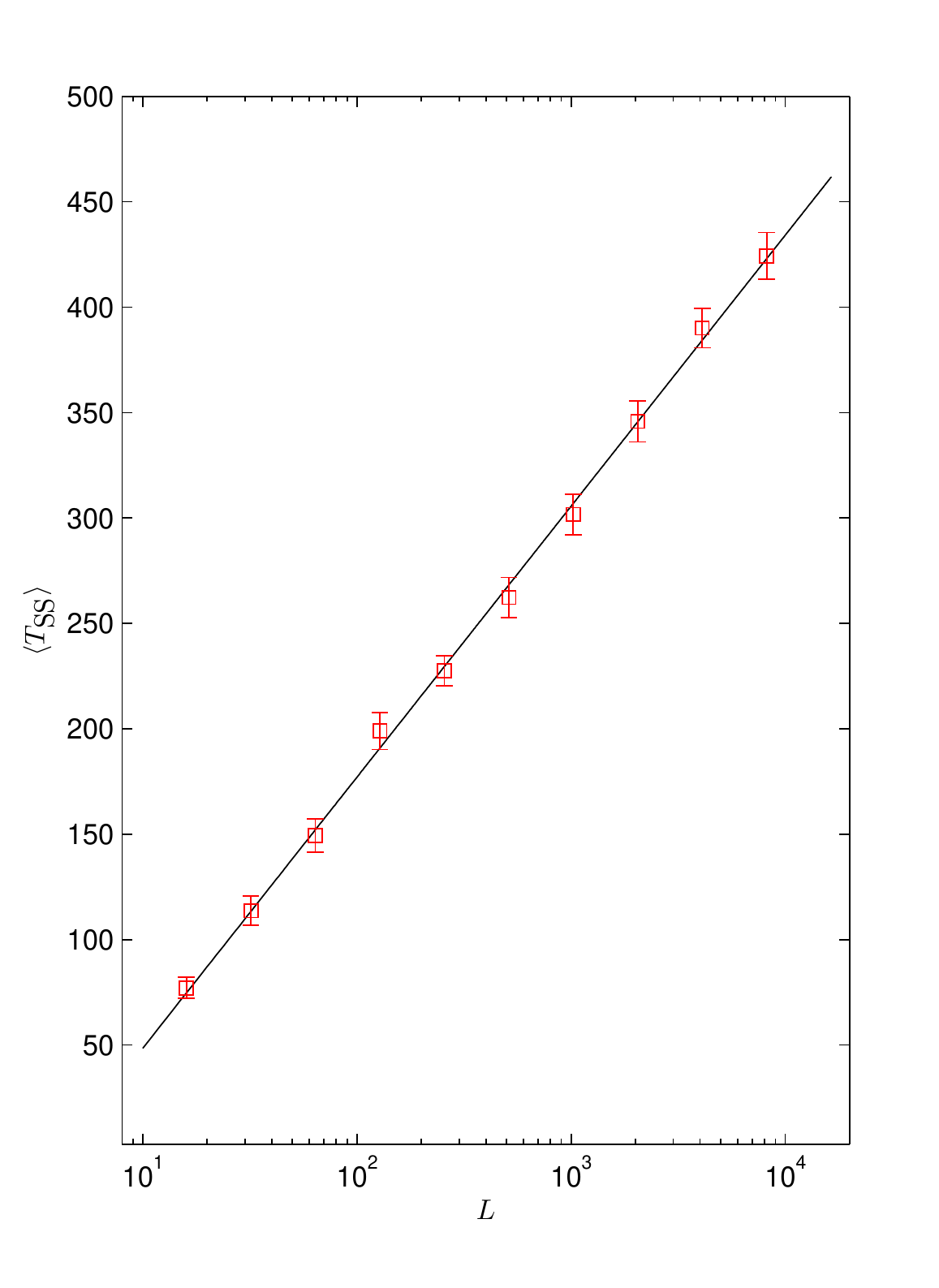}
}
\end{center}
\caption{\label{fig:exp}
The second moment $\sigma^2 (t)$ increases asymptotically exponentially with time as predicted in (\ref{mexp}) given by a full line, before it saturates at the stationary value $(\rho -\rho_c )^2 L$ indicated by dashed lines (left). Parameter values are $\gamma =3$, $d=0.1$, $\rho =2$, using the rates (\ref{warates}), and fit constants are $C_1 =25$, $C_2 =0.018$. This leads to a logarithmic increase for $\langle T_{SS}\rangle$ as shown on the right, which fits well the prediction $\log \frac{(\rho -\rho_c )^2}{C_1} +\frac{1}{C_2}\log L$ given by a full line. 
Data points are averaged over $100$ realizations and errors are comparable to the size of the symbols unless indicated.
}
\end{figure}

For the boundary case $\gamma =3$ in one dimension, our results (\ref{ccase}) and (\ref{ecase}) predict a linear equation
\be
\frac{d}{dt} m(t)=\mathrm{const.}\ m(t)\ ,
\ee
with an exponential solution for the growth of the mean cluster size. Analogously, this leads to the prediction for the second moment
\be\label{mexp}
\sigma^2 (t)=C_1 e^{C_2 t}\quad\mbox{with constants }C_1 ,C_2 >0\ .
\ee
It reaches the stationary value $(\rho -\rho_c )^2 L$ in a time of order $\log L$, which is the expected scaling of $\langle T_{SS}\rangle$ in this case. This behaviour is confirmed in Figure \ref{fig:exp} for the system with the rates (\ref{warates}) originally studied in \cite{Waclaw:2012ww}, confirming that our analysis does not depend on the detailed functional form of the rates but only on their asymptotic behaviour.

\section{Summary and discussion\label{sec:sum}}

We have established a complete picture of condensation dynamics in a symmetric version of the model introduced in \cite{Waclaw:2012ww}, which includes an explosive regime for parameter values $\gamma >3$, and a novel coarsening regime for $\gamma\in (2,3)$. The stages of the condensation dynamics can be summarized as follows:

\begin{itemize}
	\item Initialization: due to fluctuations in the initial conditions or the dynamics at a very early stage, the largest occupation numbers in the system scale as $\log L$ with the system size $L$.
	\item Nucleation: amplification of these initial fluctuations by local dynamics leads to large clusters of the critical system size $m_c \sim L^{2/(\gamma -1)}$ (\ref{mc}) within a time of order $(\log L)^{3-\gamma}$ (\ref{nutime}) for $\gamma >3$. For $\gamma\in (2,3)$ macroscopic clusters of size $O(L)$ are established in a time of order $L^{3-\gamma}$ (\ref{nutime}).
\end{itemize}
The remaining time evolution is then governed by two distinct global dynamics:
\begin{itemize}
	\item Coarsening for $\gamma\in (2,3)$: macroscopic clusters move across the lattice and exchange particles or merge on a time scale $L^{3-\gamma}$ which overlaps with the nucleation regime, leading to $\langle T_{SS}\rangle \sim L^{3-\gamma}$ (\ref{tssc}).
	\item Explosive condensation for $\gamma >3$: a single cluster dominates the process, covering the lattice in vanishing time and growing to a single condensate in an almost deterministic fashion. This is clearly separated from and much faster than the nucleation regime, which dominates the time scale $\langle T_{SS}\rangle \sim (\log L)^{3-\gamma}$ (\ref{tsse}).
\end{itemize}
For the asymmetric transport model, the threshold $\gamma>2$ for the onset of explosive condensation agrees with the threshold for instantaneous gelation in mean field theory of exchange-driven growth \cite{ben-naim_exchange-driven_2003}, whereas for symmetric transport, $\gamma>3$ is required. This is consistent with the expectation that spatial models become mean-field like in their behaviour only when the spatial transport mechanism causes sufficiently strong mixing to break spatial correlations. It is clear that symmetric transport will mix more slowly than asymmetric transport. Our analysis suggests that this difference makes possible the existence of the window $2<\gamma<3$ in which spatial correlations in the coarsening process in one dimension remain sufficiently strong
to prevent the explosive behaviour present at mean-field level.

Our theoretical arguments are slightly different from previous work on asymmetric systems \cite{Waclaw:2012ww,Evans:2014dy}, but can be directly adapted to the that situation, which leads to simpler explanations of the explosive condensation regime. Our results imply that asymmetry of the dynamics is in fact not necessary for explosive condensation, and that the phenomenon is simply caused by strong enough non-linearity of the jump rates. Our arguments can also be directly generalized to higher dimensions. It is known that the expected number of steps for a symmetric random walk to visit all sites of a $d$-dimensional lattice with side length $L$ scales like the volume $L^d$ for $d\geq 2$, with logarthmic corrections  for $d=2$ and $3$. With (\ref{mc}), the critical cluster size then scales like
\be
m_c \sim L^{d/(\gamma -1)}\ll L^d\quad\mbox{for }d\geq 2\mbox{ and }\gamma >2\ .
\ee
Therefore, we expect explosive condensation for symmetric systems in higher dimensions for all values of $\gamma >2$, resulting from the scaling of cover times for symmetric random walks being similar to asymmetric ones. The non-explosive behaviour for $\gamma \in (2,3)$ in one dimension is in fact a special case resulting from the $L^2$ scaling of cover times and recurrence of random walks. The transient behaviour of higher dimensional random walks also implies that clusters gain mass at an increased rate $D(m)=m^{\gamma -1}$, which leads to a vanishing nucleation time of order $(\log L)^{2-\gamma}$ for all $\gamma >2$ as a correction to (\ref{nutime}). This is then the expected scaling for $\langle T_{SS}\rangle$ in all higher dimensions. Numerical data of the quality to test predicted scaling laws in higher dimensions are hard to obtain and beyond the scope of this paper. The spatial part $p(x,y)$ does not have to be nearest neighbour, our results apply directly as long as it is irreducible and spatially homogeneous. A further 
interesting question in this context is the role of inhomogeneities, and in particular whether the model could be non-explosive on (random) graphs which have bottlenecks and large cover times for random walks.

In general, the simulation of particle systems with non-linearly increasing jump rates is computationally very demanding. Rejection based algorithms such as random sequential update are highly inefficient, and the only option is Gillespie's algorithm which generates the sample path of general continuous-time Markov chains \cite{gillespie}. A single step of a macroscopic cluster requires of order $L^2$ individual transitions of particles due to symmetry of the dynamics. This is a factor $L$ more than for asymmetric models covered in previous works, which explains the slightly poorer quality of our numerical results. Choosing small parameter values for $d$ decreases the critical density and leads to a clearer separation between the condensed and the fluid phase. This improves the quality of numerical data and also their agreement with our theoretical arguments which are based on this separation, which is evident from Figures \ref{fig:tss} and \ref{fig:tss2}. It also allows us to replace the full dynamics of a 
cluster step by an effective single transition in the late stages of the dynamics, which is particularly efficient in the regime $\gamma >3$. Details on this can be found in \cite{yuxi}, Chapter 5.

\section*{Acknowledgements}
C.C. and S.G. acknowledge support by the Engineering and Physical Sciences Research Council (EPSRC), Grant Nos. EP/I014799/1 and EP/I01358X/1.

\section*{References}


\end{document}